\documentclass[a4paper,american,prl,superscriptaddress,longbibliography,fixfloat,notitlepage]{revtex4-1}
\usepackage[latin9]{inputenc}
\usepackage{color}
\usepackage{amsmath}
\usepackage{amssymb}
\usepackage{graphicx}

\makeatletter

\pdfpageheight\paperheight
\pdfpagewidth\paperwidth

\usepackage{xcolor}
\definecolor{darkblue}{rgb}{0.1,0.2,0.6} 
\definecolor{lightblue}{rgb}{0.1,0.1,1.0}
\definecolor{darkred}{rgb}{0.8,0.1,0.2}
\usepackage[colorlinks,citecolor=lightblue,linkcolor=darkblue,urlcolor=lightblue] {hyperref}

\makeatother

\usepackage{babel}
\begin{document}
\global\long\def\E{\mathrm{e}}%
\global\long\def\D{\mathrm{d}}%
\global\long\def\I{\mathrm{i}}%
\global\long\def\mat#1{\mathsf{#1}}%
\global\long\def\vec#1{\mathsf{#1}}%
\global\long\def\cf{\textit{cf.}}%
\global\long\def\ie{\textit{i.e.}}%
\global\long\def\eg{\textit{e.g.}}%
\global\long\def\vs{\textit{vs.}}%
 
\global\long\def\ket#1{\left|#1\right\rangle }%
 
\global\long\def\etal{\textit{et al.}}%
\global\long\def\tr{\text{Tr}\,}%
 
\global\long\def\im{\text{Im}\,}%
 
\global\long\def\re{\text{Re}\,}%
 
\global\long\def\bra#1{\left\langle #1\right|}%
 
\global\long\def\braket#1#2{\left.\left\langle #1\right|#2\right\rangle }%
 
\global\long\def\obracket#1#2#3{\left\langle #1\right|#2\left|#3\right\rangle }%
 
\global\long\def\proj#1#2{\left.\left.\left|#1\right\rangle \right\langle #2\right|}%

\title{How many particles make up a chaotic many-body quantum system?}
\author{Guy Zisling}
\affiliation{Department of Physics, Ben-Gurion University of the Negev, Beer-Sheva
84105, Israel}
\author{Lea F. Santos}
\affiliation{Department of Physics, Yeshiva University, New York, New York 10016,
USA}
\author{Yevgeny Bar Lev}
\affiliation{Department of Physics, Ben-Gurion University of the Negev, Beer-Sheva
84105, Israel}
\email{ybarlev@bgu.ac.il}

\begin{abstract}
We numerically investigate the minimum number of interacting particles,
which is required for the onset of strong chaos in quantum systems
on a one-dimensional lattice with short-range and long-range interactions.
We consider multiple system sizes which are at least three times larger
than the number of particles and find that robust signatures of quantum
chaos emerge for as few as 4 particles in the case of short-range
interactions and as few as 3 particles for long-range interactions,
and without any apparent dependence on the size of the system.
\end{abstract}
\maketitle

\section{Introduction}

Quantum chaos, especially when caused by particle interactions, has
seen a revival in the last decade or so, because it is closely related
with topics of high experimental and theoretical interest. It is behind
the mechanism of thermalization of isolated many-body quantum systems
and the validity of the eigenstate thermalization hypothesis (ETH)
\citep{Borgonovi2016,Dalessio2016,Kaufman2016}, it explains the heating
of driven systems~\citep{Dalessio2014,lazarides2015fate}, it is
the main obstacle for many-body localization~\citep{Schreiber2015,Nandkishore2015,Luitz2017,suntajs2020quantum},
it inhibits long-time simulation of many-body quantum systems~\citep{Yang2020},
it can lead to the fast scrambling of quantum information~\citep{Sanchez2020},
and it is the regime where the phenomenon of quantum scarring may
be observed~\citep{Bernien2017,Turner2018,pilatowskycameo2020does}.

For systems with a proper semiclassical limit, quantum chaos refers
to specific properties found in the quantum domain, when the corresponding
classical system is chaotic in the sense of mixing, sensitivity to
initial conditions and positive Lyapunov exponents. This correspondence
is well established for systems with a few degrees of freedom, such
as billiards and kicked rotors, however in the case of systems with
many interacting particles, as the ones we are interested in, the
correspondence is still lacking due to the challenges involved in
their semiclassical analysis~\citep{Akila2017}. The usual approach
is therefore to denote a given system as chaotic if it shows correlated
eigenvalues and eigenstates components with similar features to those
found in ensembles of full random matrices~\citep{Guhr1998,StockmannBook,Flambaum1994,Zelevinsky1996}.

Most recent studies of quantum chaos in many-body systems are performed
for a finite density of particles, but two questions arise: \emph{can
quantum chaos occur also at the limit of zero density?} And if so,
\emph{how many interacting particles are needed to bring a quantum
system to the regime of strong chaos?} These questions are particularly
relevant for experiments with cold atoms and ion traps, where the
number of particles and also the size of the systems can be controlled.
In Ref.~\citep{Wenz2013}, by increasing the number of cold atoms
step by step, it was experimentally shown that the Fermi sea is formed
for as few as four particles. Quantum chaos \citep{Flambaum1994}
and thermalization with the appearance of the Fermi-Dirac distribution
\citep{Schnack1996,Schnack2000,Flambaum1996b,PhysRevE.56.5144,Izrailev2001}
were also obtained with just four interacting particles. More recently,
thermalization was studied in systems with 5 particles \citep{Rigol2008}
and quantum chaos was verified again in systems with only 4 particles
\citep{Harshman2017,Schiulaz2018,Fortes_2020,PhysRevE.100.042201},
and possibly even with as few as 3 interacting particles \citep{mirkin2020small}However,
it is not entirely clear if other indicators of chaos show similar
behaviors, and if the obtained threshold of 4 interacting particles
can be changed by the introduction of long-range interactions. These
are the questions that we consider in this work.

We focus on spin-1/2 chains with a small number $N$ of excitations
and power-law interactions that decay with the distance between the
spins. These systems are analogous to systems of hardcore bosons or
spinless fermions, such that the number of particles in these cases
corresponds to the spin excitations in our models \footnote{The analogy is exact only in the limit of nearest-neighbors interactions,
as can be seen via the Jordan-Wigner and Holstein-Primakoff transformations}. We find that in systems with short-range couplings, strong chaos
emerges already for $N\gtrsim4$, no matter how large the system size
is. While large chains improve the statistics, they do not change
our results. We show that long-range interactions can facilitate the
transition to chaos and decrease the threshold to only 3 excitations,
such that systems with only 3 interacting particles exhibit chaotic
properties similar to large interacting systems in the dense limit.
This is of particular interest to experiments with ion traps, where
the range of interactions can be controlled~\citep{Jurcevic2014,Richerme2014},
and to studies which explore the generalization of the Lieb-Robinson
bound for long-range interacting systems~\citep{Hauke2013,Jurcevic2014,Richerme2014,SantosCelardo2016}.

\section{Model and Chaos Indicators}

The spin-1/2 chain that we study is described by the following Hamiltonian
\begin{equation}
\hat{H}_{\gamma}=\sum_{i=1}^{L-1}\sum_{j=i+1}^{L}\frac{J}{\left(j-i\right)^{\gamma}}\left(\hat{S}_{i}^{x}\hat{S}_{j}^{x}+\hat{S}_{i}^{y}\hat{S}_{j}^{y}+\Delta\hat{S}_{i}^{z}\hat{S}_{j}^{z}\right)+h_{1}\hat{S}_{1}^{z}+h_{\lfloor L/2\rfloor}\hat{S}_{\lfloor L/2\rfloor}^{z},\label{eq:long xxz}
\end{equation}
where $\hat{S}_{i}^{x,y,z}$ are the spin-1/2 operators at lattice
site $i$, $L$ is the size of the chain, $J$ is the coupling strength,
which we set to be equal to 1, $\Delta$ is the anisotropy parameter,
$\gamma$ determines the range of the interactions, whose strengths
decay as a power-law with the distance between the spins; and $h_{1}$
{[}$h_{\lfloor L/2\rfloor}${]} is the amplitude of an impurity (defect)
placed on site $1$ {[}site $\lfloor L/2\rfloor${]}. The system has
open boundary conditions and conserves the total magnetization in
the $z$-direction.

We analyze the onset of chaos in sectors with low total $z$-magnetization,
where most spins, except a few, point down. Each such sector is characterized
by the number of up-spins (excitations). In equivalent models of spinless
fermions, this number corresponds to the number of particles. We denote
the number of excitations by $N$ and the Hilbert space dimension
of the corresponding sector by $\mathcal{D}={L \choose N}$.

\subsection{Integrable and chaotic points}

In the limit of $\gamma\to\infty$ and for $h_{1}=h_{\lfloor L/2\rfloor}=0$,
Eq.~(\ref{eq:long xxz}) describes the XXZ model with nearest-neighbor
couplings, which is an integrable model. By adding a small impurity
at the border of the chain, we can break the reflection symmetry (parity),
but the model remains integrable. Throughout this work we fix $h_{1}=0.11$
and denote this integrable point by $\hat{H}_{\infty}^{\text{0}}$,
where the superscript indicates that $h_{\lfloor L/2\rfloor}=0$.
To avoid degeneracies, we stay away from the isotropic point and fix
$\Delta=0.55$.

We explore two ways to break the integrability of the XXZ model: by
adding an impurity in the middle of the chain, $h_{\lfloor L/2\rfloor}\neq0$,
and by adding long-range interactions. The fact that the addition
of a defect takes the system to the chaotic regime was demonstrated
in Refs.~\citep{Santos2004,Torres2014PRE,Santos2020,TorresKollmar2015}.
Here, without the loss of generality, we choose $h_{\lfloor L/2\rfloor}=0.7$
and denote the single-impurity model with nearest-neighbor interactions
by $\hat{H}_{\infty}^{\text{imp}}$. In the absence of the middle-site
impurity, integrability is broken by adding interactions between further
neighbors, which we do by decreasing the value of $\gamma$. For $h_{\lfloor L/2\rfloor}=0$,
the system approaches the chaotic domain for $\gamma\lessapprox5$,
but then gets closer to yet another integrable point for $\gamma<1$.
We therefore focus on the interval $1\leq\gamma\leq5$.

\subsection{Indicators of chaos}

We employ two indicators of quantum chaos that do not require the
unfolding of the spectrum. To detect short-range correlations between
the eigenvalues, we use the so-called r-metric, which was introduced
in Refs.~\citep{Oganesyan2007,Atas2012,Corps2020},
\begin{equation}
r_{\alpha}=\min\left(\frac{s_{\alpha}}{s_{\alpha-1}},\frac{s_{\alpha-1}}{s_{\alpha}}\right),
\end{equation}
where $s_{\alpha}=E_{\alpha+1}-E_{\alpha}$ is the spacing between
neighboring eigenvalues of the Hamiltonian. Averaging over all the
eigenvalues, $\langle r\rangle\approx0.39$ for the Poissonian distribution
of the spacings, that is often found in integrable models. For chaotic
models with real and symmetric matrices, $\langle r\rangle\approx0.536$.
While in our calculations of $\langle r\rangle$, we consider the
whole spectrum, it is worth emphasizing that in realistic systems,
as the ones we study, chaos develops away from the edges of the spectrum.

Since the eigenstate thermalization hypothesis (ETH) holds due to
quantum chaos, we can use the indicators of ETH to detect the transition
to chaos. The expectation value of an observable $\hat{O}$ evolves
according to 
\begin{equation}
O(t)=\left\langle \Psi\left|\hat{O}\left(t\right)\right|\Psi\right\rangle =\sum_{\alpha}\left|C_{\alpha}\right|{}^{2}O_{\alpha\alpha}+\sum_{\alpha\neq\beta}C_{\alpha}^{*}C_{\beta}e^{-i(E_{\beta}-E_{\alpha})t}O_{\alpha\beta},\label{Eq:O}
\end{equation}
where $C_{\alpha}=\langle\alpha|\Psi\rangle$ is the overlap between
the eigenstate $|\alpha\rangle$ and the initial state $|\Psi\rangle$
of the system and $O_{\alpha\beta}=\langle\alpha|\hat{O}|\beta\rangle$.
For sufficiently local observables, ETH builds on two assumptions:
that the infinite-time average of $O\left(t\right)$, which corresponds
to the first term in Eq.~(\ref{Eq:O}), coincides with the value
of the operator at thermal equilibrium, and that the fluctuations
around this value, which are given by the second term in Eq.~(\ref{Eq:O}),
decrease with system size and cancel out on average. In this work,
we focus on the second term, in particular, we investigate the distributions
of the off-diagonal elements of the operator, $O_{\alpha\beta}$.

The distribution of the off-diagonal matrix elements, $O_{\alpha\beta}$,
in chaotic (thermalizing) systems is Gaussian~\citep{Beugeling2015,LeBlond2019,Brenes2020,richter2020eigenstate,PhysRevB.102.075127,schonle2020eigenstate},
while integrable models have a clear non-Gaussian distribution~\citep{feingold1986distribution,LeBlond2019,luitz2016anomalous,PhysRevB.102.075127}.
The observable that we consider is the magnetization on the impurity-site,
$O_{\alpha\beta}=\left\langle \alpha\left|\hat{S}_{\lfloor L/2\rfloor}^{z}\right|\beta\right\rangle $,
and to assess the chaoticity of the studied systems, we quantify the
distance of the distribution of $O_{\alpha\beta}$ from a normal distribution
using two measures.

One quantity considered is the kurtosis of the distribution of $O_{\alpha\beta}$,
\begin{equation}
\kappa_{\hat{O}}=\frac{1}{\sigma^{4}}\left\langle \left(O_{\alpha\beta}-\left\langle O_{\alpha\beta}\right\rangle \right){}^{4}\right\rangle ,\label{eq:kurtosis}
\end{equation}
where $\langle.\rangle$ indicates the average over all pairs of eigenstates
$|\alpha\rangle\neq|\beta\rangle$ and $\sigma$ is the standard deviation
of the distribution of $O_{\alpha\beta}$. For Gaussian distributions
the kurtosis is $\kappa_{\hat{O}}=3$. In our plots of the distributions
and in our calculations of $\kappa$, we always consider 200 eigenstates
with energies closest to the center of the many-body spectrum.

The other metric that we use is 
\begin{equation}
\Gamma_{\hat{O}}\left(\omega=E_{\beta}-E_{\alpha}\right)=\frac{\overline{\left|O_{\alpha\beta}\right|^{2}}}{\overline{\left|O_{\alpha\beta}\right|}^{2}},\label{eq:Gamma}
\end{equation}
which allows to assess the departure from the Gaussianity of the distribution
as a function of the energy difference $\omega=E_{\beta}-E_{\alpha}$.
We extract all the eigenstates that satisfy $(E_{\alpha}+E_{\beta})/2\in[-0.025\epsilon,+0.025\epsilon]$,
where $\epsilon$ is the many-body bandwidth, $\epsilon\equiv E_{max}-E_{min}$,
and group these pairs according to their value of $\omega$ in bins
of width 0.05 . The overbar in Eq.~(\ref{eq:Gamma}) indicates averaging
over the pairs in a given bin. For a Gaussian distribution the value
of $\Gamma_{\hat{O}}$ does not depend on $\omega$ and is equal to
$\pi/2$ \citep{Geary1935}.

\section{Results}

We now have all the tools to investigate how the transition to chaos
depends on the number of excitations, $N,$ for systems with short
and long-range interactions.

\subsection{Short-Range Interactions}

\begin{figure*}[t]
\includegraphics{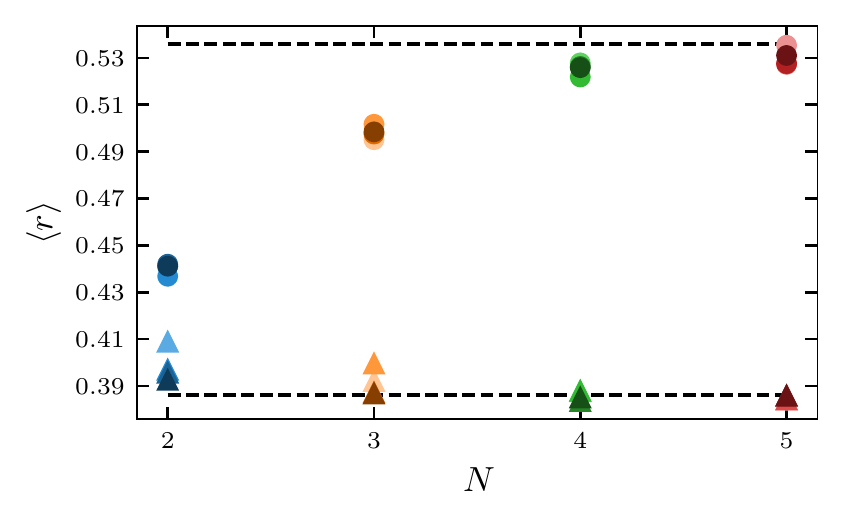} \hspace{0.1cm}
\includegraphics{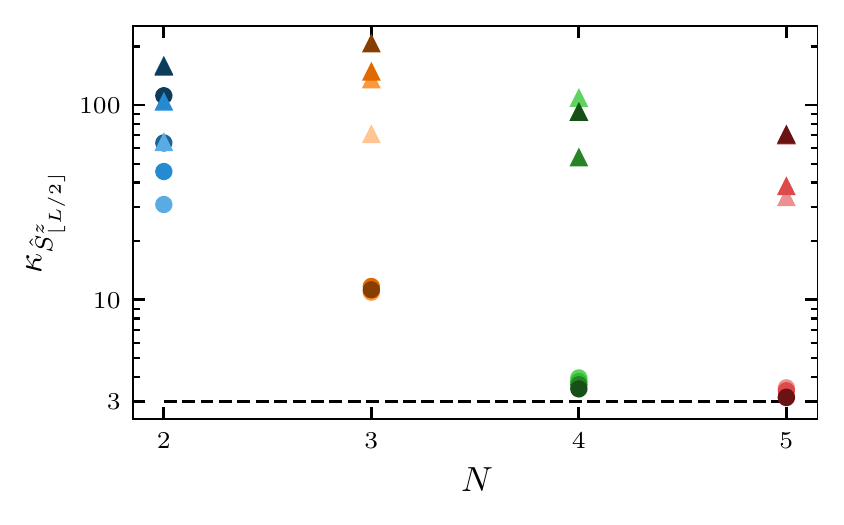} \caption{\emph{Left panel}: Quantum chaos indicator $\left\langle r\right\rangle $
as a function of the number of excitations $N$ calculated for the
integrable model $\hat{H}_{\infty}^{0}$ ($\blacktriangle$) and the
single-impurity model $\hat{H}_{\infty}^{\text{imp}}$ ($\text{\ensuremath{\bullet}}$).
\emph{Right panel}: same as the left panel, but for the kurtosis $\kappa$
of the distribution of the off-diagonal elements of the middle-site
magnetization. Each color represents a number of excitations with
the darker shades corresponding to larger system sizes. The system
sizes ranges used are: $N=2,$$L\in\left[100,200\right]$; $N=3,$$L\in\left[30,50\right]$;
$N=4,$$L\in\left[22,28\right]$ and $N=5,$$L\in\left[16,21\right]$.}
\label{fig:short_r_metric}
\end{figure*}

We start our analysis by considering the limit of short-range couplings,
$\gamma\to\infty$. In the left panel of Fig.~\ref{fig:short_r_metric},
we plot $\left\langle r\right\rangle $ as a function of the number
of excitations for various systems sizes for $\hat{H}_{\infty}^{0}$
\textcolor{black}{(triangles)} and $\hat{H}_{\infty}^{\text{imp}}$
(circles). As expected, for $\hat{H}_{\infty}^{0}$, $\left\langle r\right\rangle $
stays very close to 0.39 indicating integrability, with negligible
drifts with the system size. On the other hand, for $\hat{H}_{\infty}^{\text{imp}}$,
the metric $\left\langle r\right\rangle $ increases gradually from
an intermediate value between integrability and chaos, $\langle r\rangle\simeq0.44$,
obtained for two excitations, to the chaotic value of $\langle r\rangle\simeq0.536$
for four or more excitations. The size of the chain does not affect
the results.

\begin{figure*}[h!]
\includegraphics{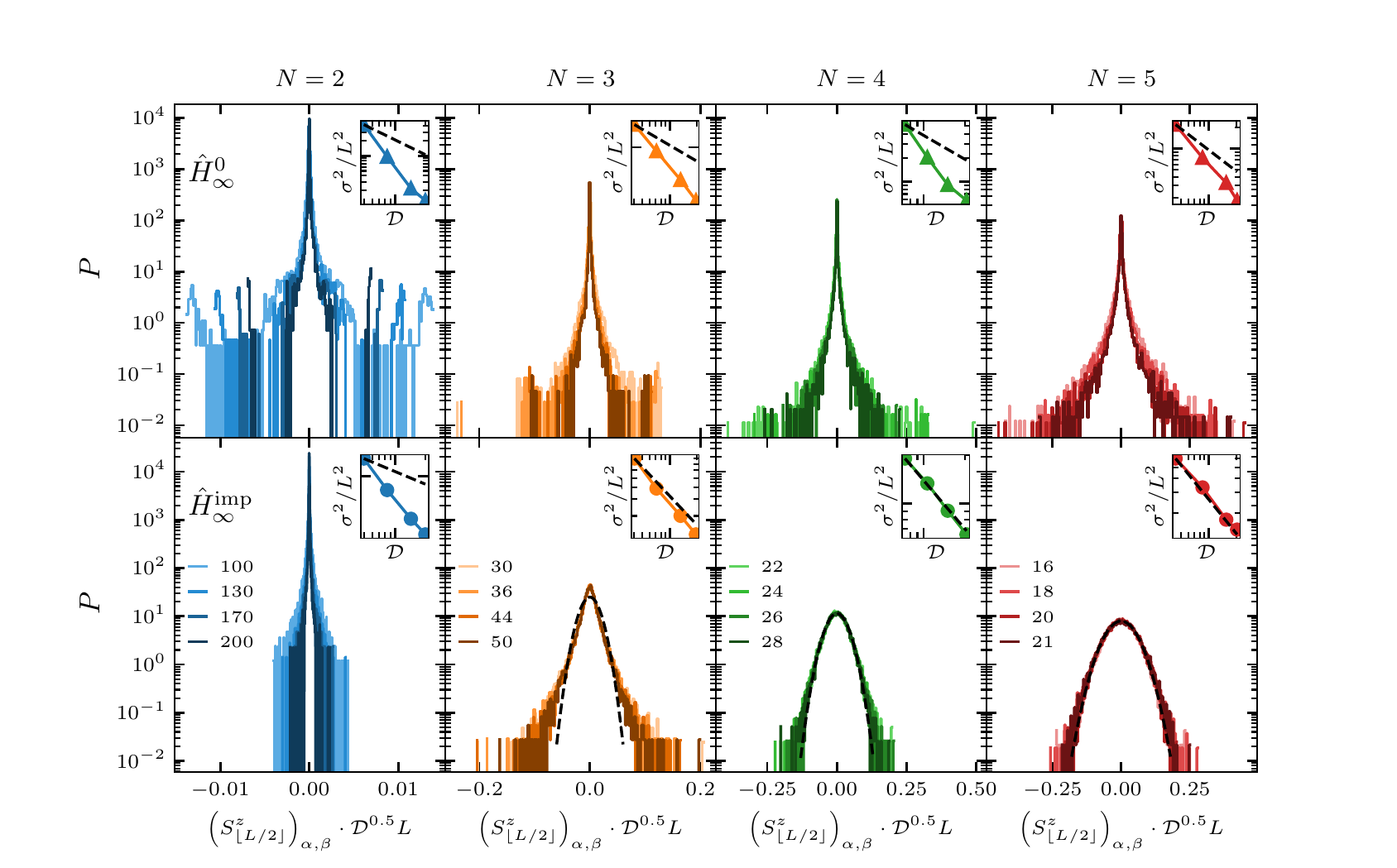} \caption{Distribution of the off-diagonal elements of $\hat{S}_{\lfloor L/2\rfloor}^{z}$,
computed for 200 eigenstates in the middle of the spectrum, for the
integrable model $\hat{H}_{\infty}^{0}$ (top row) and the single-impurity
model $\hat{H}_{\infty}^{\text{imp}}$ (bottom row) for different
number of excitations (different columns). In each panel, larger system
sizes are represented by darker colors (see legends). The histograms
are scaled by $\mathcal{D}^{0.5}L$. The insets display log-log plots
of the scaled variance $\sigma^{2}/L^{2}$ against the Hilbert space
dimension and the black dashed lines correspond to $\sigma^{2}/L^{2}\propto\mathcal{D}^{-1}$.}
\label{fig:offdiag_dist fze short range model}
\end{figure*}

Figure~\ref{fig:offdiag_dist fze short range model} shows the distributions
of the off-diagonal elements of $\hat{S}_{\lfloor L/2\rfloor}^{z}$.
For chaotic systems without conserved quantities the variance of the
off-diagonal matrix elements scales as $\mathcal{D}^{-1},$where $\mathcal{D}$
is the Hilbert space dimension. However when conserved quantities
are present, the variance decreases slower, as $L^{2}\mathcal{D}^{-1}$\citep{Dalessio2016,luitz2016anomalous}.
Since in our case both the energy and the magnetization are conserved,
to plot the distributions of the off-diagonal matrix elements corresponding
to different system sizes, such that they will have the same variance,
we rescale the values of the off-diagonal matrix elements by the factor
$\mathcal{D}^{0.5}L$.

For $\hat{H}_{\infty}^{0}$ (top row in Fig.~\ref{fig:offdiag_dist fze short range model}),
the distributions are visibly non-Gaussian and exhibit a peaked structure
for any number of excitations. For $\hat{H}_{\infty}^{\text{imp}}$
(bottom row in Fig.~\ref{fig:offdiag_dist fze short range model})
they are non-Gaussian for $N=2,3$, but this changes for $N\geq4$,
which is consistent with our results for $\left\langle r\right\rangle $
in the left panel of Fig.~\ref{fig:short_r_metric}, where the single-impurity
model shows a transition to the regime of strong chaos for 4 or more
excitations. For $N\geq4$ the variance of the off-diagonal matrix
elements, as seen in the insets of Fig.~\ref{fig:offdiag_dist fze short range model},
decreases as $L^{2}\mathcal{D}^{-1},$ as expected for chaotic systems
with conserved quantities \citep{Dalessio2016,luitz2016anomalous}.
Notice also that scaled distributions do not show significant dependence
on system size, although due to better statistics, the curves become
smoother as $L$ increases.

To quantify how close the distributions are to normal distributions,
we plot their kurtosis in the right panel of Fig.~\ref{fig:short_r_metric}.
For $\hat{H}_{\infty}^{0}$ \textcolor{black}{(triangles)}, the kurtosis
is much larger than the value which corresponds to a normal distribution,
$\kappa=3$, and it increases with $L$. For $\hat{H}_{\infty}^{\text{imp}}$
(circles) the kurtosis is close to 3 for $N\geq4$, converging even
closer to 3 as the system size increases.

The behavior of $\left\langle r\right\rangle $ in the left panel
of Fig.~\ref{fig:short_r_metric} and of the kurtosis in the right
panel of Fig.~\ref{fig:short_r_metric} for $\hat{H}_{\infty}^{\text{imp}}$
shows a very similar trend towards chaos as $N$ increases, namely,
as $\left\langle r\right\rangle $ approaches its chaotic value 0.536,
$\kappa$ approaches 3. We note that these two metrics are very different
in nature, since the r-metric has information about the spectrum,
while the kurtosis reflects the structure of the eigenstates through
the off-diagonal elements of the observable, yet, they provide equivalent
information about the onset of quantum chaos.

\begin{figure*}[h]
\includegraphics{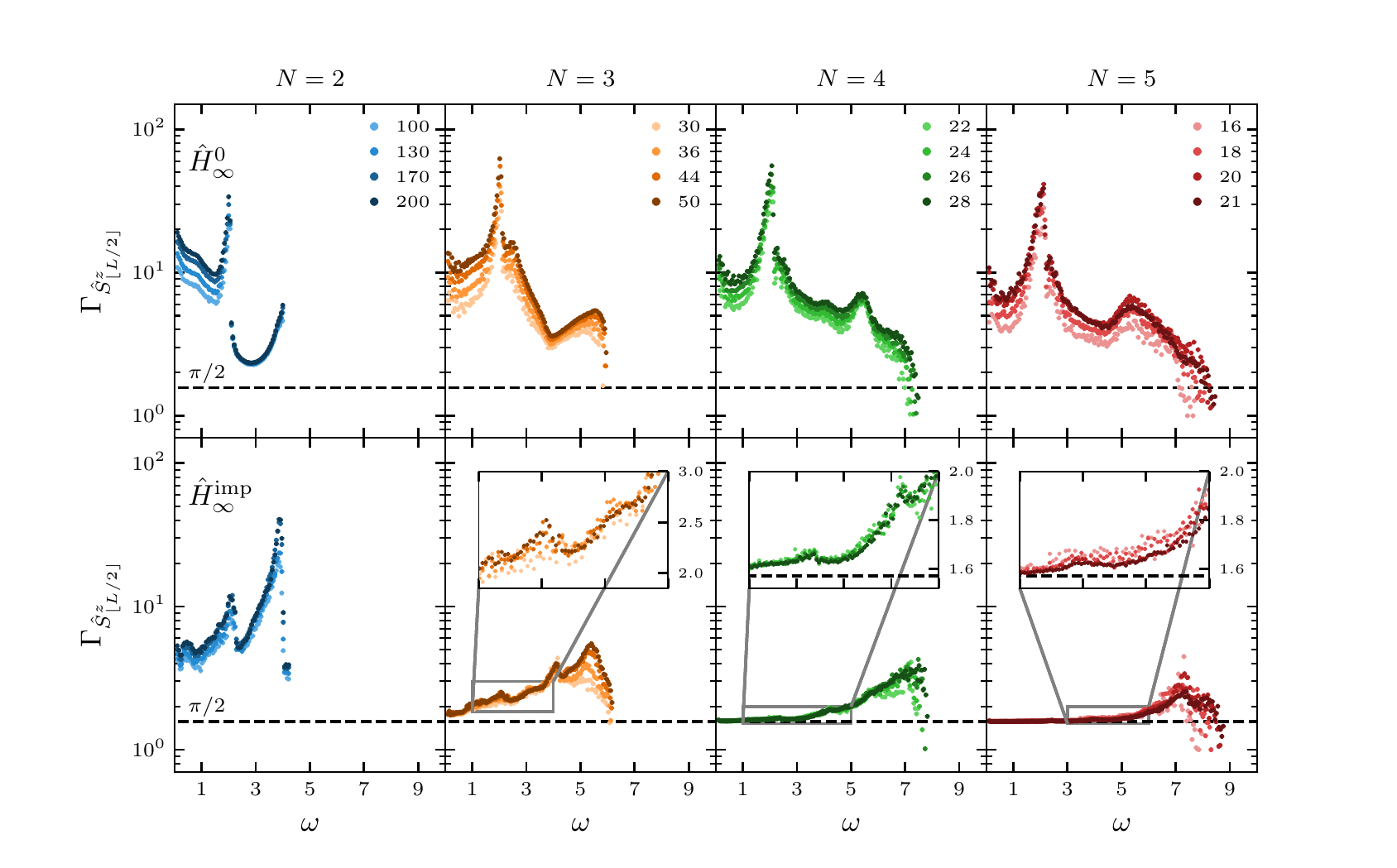} \caption{Ratios $\Gamma_{\hat{S}_{\lfloor L/2\rfloor}^{z}}\left(\omega\right)$
for the integrable model $\hat{H}_{\infty}^{0}$ (top row) and the
single-impurity model $\hat{H}_{\infty}^{\text{imp}}$ (bottom row)
for different number of excitations (different columns). In each panel,
larger system sizes are represented by darker colors (see legends).
The insets zoom in into areas of interest and are on a linear scale.
See main text for the explanation of how $\Gamma_{\hat{S}_{\lfloor L/2\rfloor}^{z}}\left(\omega\right)$
was calculated.}
\label{fig:short gamma}
\end{figure*}

In Fig.~\ref{fig:short gamma}, we analyze how close the distributions
of the off-diagonal elements of $\hat{S}_{\lfloor L/2\rfloor}^{z}$
are to normal distributions, taking into account the energy difference
$\omega$. For $\hat{H}_{\infty}^{0}$ (top row), the value of $\Gamma_{\hat{S}_{\lfloor L/2\rfloor}^{z}}$
is larger than the value which corresponds to a normal distribution,
$\Gamma_{\hat{S}_{\lfloor L/2\rfloor}^{z}}=\pi/2,$ for all $\omega$
and it increases as the system size increases. The behavior is very
similar for $\hat{H}_{\infty}^{\text{imp}}$ and $N\leq3$, although
for $N=3$, as the inset indicates, the deterioration with the system
size is less apparent. For $\hat{H}_{\infty}^{\text{imp}}$ and $N=4$,
we reach a crossing point, where $\Gamma_{\hat{S}_{\lfloor L/2\rfloor}^{z}}$
is close to $\pi/2$ for small $\omega$'s and appears to converge
to $\pi/2$ with the system size. The improvement with $L$ for $N=5$
is even more evident and somewhat analogous to the improvement with
system size verified in systems with a fixed density \citep{LeBlond2019,Brenes2020,PhysRevB.102.075127}.

\subsection{Long-Range Interactions}

\begin{figure*}
\includegraphics{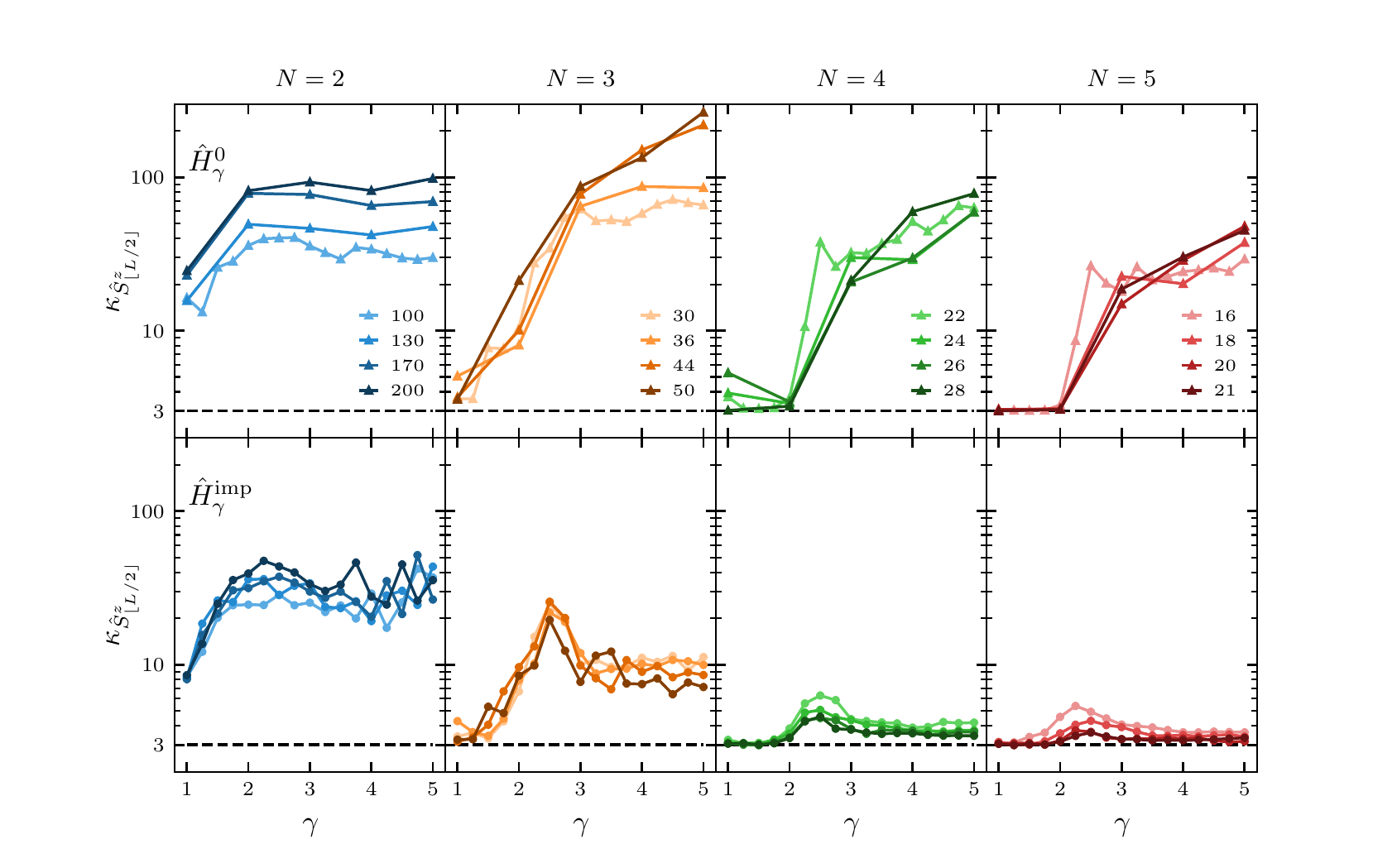} \caption{Kurtosis of the distribution of off-diagonal matrix elements computed
for 200 eigenstates in the middle of the spectrum for systems with
variable ranges of interactions without the middle-site impurity,
$\hat{H}_{\gamma}^{0}$ (top row), and with a middle-site impurity,
$\hat{H}_{\gamma}^{\text{imp}}$ (bottom row), plotted for different
number of excitations (different columns). In each panel, larger system
sizes are represented by darker shades (see legends).}
\label{fig:kurt-vs-gamma-mid-def-fze}
\end{figure*}

We now examine how the transition to quantum chaos is affected by
the the presence of long-range interactions. Given the similar information
obtained with the spectral correlation measure $\left\langle r\right\rangle $
and the kurtosis, in Fig.~\ref{fig:kurt-vs-gamma-mid-def-fze} we
present only the kurtosis as a function of the coupling range $\gamma$.

For long-range interactions, when $\gamma\sim1$, the system approaches
the chaotic limit, corresponding to $\kappa=3$, for as few as 3 excitations
and this happens for both $\hat{H}_{\gamma}^{0}$ (top row) and $\hat{H}_{\gamma}^{\text{imp}}$
(bottom row). Focusing on the point $\gamma=1$ there is however no
apparent drift towards chaos with the system size.

For $1\leq\gamma\leq2$ and $N\geq4$, it is evident that the system
is chaotic and drifts towards chaos as a function of the system size
for both $\hat{H}_{\gamma}^{0}$ and $\hat{H}_{\gamma}^{\text{imp}}$.
As we leave the region of long-range interactions and $\gamma>2$,
the results for the two Hamiltonians become different, as can be anticipated,
since in this limit the interaction is sufficiently short-ranged that
the models become practically indistinguishable from their the local
variants (cf. right panel of Fig.~\ref{fig:short_r_metric}).\newpage{}

\section{Discussion}

In this work we have numerically studied the quantum chaotic properties
of the spectrum and the eigenstates of a prototypical spin-1/2 chain
with short and long-range interactions in the limit of a small number
of spin excitations. These systems correspond to bosonic or fermionic
systems with a small number of particles. Our focus is in the region
of the spectrum where quantum chaos is known to develop in systems
with many interacting particles, that is, away from the spectrum edges.

We have shown that a large one-dimensional lattice with only four
nearest-neighbor-interacting particles or even just three long-range-interacting
particles exhibits the same properties of quantum chaos observed in
systems with a finite density of interacting particles. Since our
results do not appear to depend on the system size, they suggest that
the transition to chaos occurs at zero particle density, though further
studies are in place, since for four or more excitations, it is challenging
to go below a density of 1/6 as the Hilbert space size gets too big
for exact diagonalization. Our result is of practical advantage for
experiments that have a control over the number of particles and the
range of interactions, such as those with ion traps, and which study
thermalization and other consequences of many-body quantum chaos.
Moreover it offers a simplified scenario for the development of semiclassical
analysis of interacting quantum systems.

A natural extension of our work is to search for the differences between
interacting chaotic systems at high particle density from those with
low and in particular zero density. A specific direction to be considered
is the effects of particle statistics, since it has marginal effects
for low densities, but not so in the high density limit. Other topics
worth investigating include the speed of the evolution, specially,
short-time dynamics, where spectral correlations are not yet relevant,
and transport behavior. These studies may reveal differences between
systems with few and many interacting particles \citep{Schiulaz2018},
which show similar level statistics and ETH indicators.
\begin{acknowledgments}
This research was supported by a grant from the United States-Israel
Binational Foundation (BSF, Grant No. 2019644), Jerusalem, Israel,
and the United States National Science Foundation (NSF, Grant No.~DMR-1936006),
and by the Israel Science Foundation (grants No. 527/19 and 218/19).
\end{acknowledgments}

\bibliography{biblio}

\end{document}